\documentclass[prb,twocolumn,amsmath,showpacs,preprintnumbers,superscriptaddress,amssymb,space]{revtex4}
\usepackage{amssymb}
\usepackage{makeidx}
\usepackage{amsmath}
\usepackage{dcolumn}
\usepackage{bm}
\usepackage{graphicx}
\usepackage{amsfonts}

\begin{document}

\title{Quantum metamaterial without local control}
\date{\today}

\author{A. Shvetsov}
\email{alexshdze@mail.ru}
\author{A. M. Satanin}
\affiliation{Nizhny Novgorod State University, Gagarin Ave., 23,
603950, Nizhny Novgorod, Russia}
\author{Franco Nori} \affiliation{Department of Physics, University of
Michigan, Ann Arbor, MI 48109-1040, USA}
\affiliation{Advanced Science Institute, RIKEN, Wako-shi, Saitama,
351-0198, Japan}
\author{S. Savel'ev}
\author{A.M. Zagoskin}
\email{a.zagoskin@lboro.ac.uk}
\affiliation{Advanced
Science Institute, RIKEN, Wako-shi, Saitama, 351-0198, Japan}
\affiliation{Department of Physics, Loughborough University,
Loughborough LE11 3TU, United Kingdom}

\begin{abstract}
A quantum metamaterial can be implemented as a quantum coherent 1D
array of qubits placed in a transmission line. The properties of
quantum metamaterials are determined by the local quantum state of
the system. Here we show that a spatially-periodic quantum state
of such a system can be realized without direct control of the
constituent qubits, by their interaction with the initializing
(``priming") pulses sent through the system in opposite directions.
The properties of the resulting quantum photonic crystal are
determined by the choice of the priming pulses. This proposal can
be readily generalized to other implementations of quantum
metamaterials.
\end{abstract}

\pacs{81.05.Xj,74.81.Fa}

\keywords{Metamaterials, Josephson junction arrays}

\maketitle

Since the pioneering papers ~\cite{1,pendry,shelby}, the
investigation of metamaterials (quasicontinuous media built of
artificial unit elements, that modify the properties of
propagating electromagnetic waves) was driven by their unusual
optical properties~\cite{review1,review2} in a wide range of frequencies. They are used to produce antireflection
coating and optical devices such as adaptive lenses, reconfigurable
mirrors, converters, etc. Another direction of research is related
to the development of metamaterials based on active units, e.g.,
Josephson junctions ~\cite{ricci,du}. The emission of terahertz
radiation due to the ac Josephson effect allows to build
Josephson-based emitters, filters, detectors and waveguides
operating in this spectral range important for applications
~\cite{3,4}. The nonlinear effective inductance of Josephson
junctions makes them a convenient basis for nonlinear
metamaterials ~\cite{5}.

Another fundamentally novel class of artificial media are
{\em quantum metamaterials}, i.e., optical media, which maintain global
quantum coherence over times exceeding the signal transition time
and allow local control over quantum states and basic properties
of their constituent elements  \cite{6}. The term reflects the fact that, similarly to
classical metamaterials, these systems allow additional ways to control the propagation of electromagnetic fields, not available to standard materials. The optical properties of the material are determined by its controllable coherent quantum dynamics.

Recently various schemes of qubit-array control have been proposed~\cite{Nori1,Nori2,Nori3}. Quantum metamaterials were
introduced in~\cite{6} as an array of superconducting charge qubits placed
inside a transmission line; the extensions to other
implementations, including those in optical range, soon followed
\cite{ZPSS,Zh,Fel,Haviland2011,ZagoskinBook,NoriReviews,Q,Zueco2012,ZOP}. In particular,
experiments on a flux qubit inside a transmission line reproduce
a number of atomic spectroscopy effects \cite{A2}.
Aside from the specific properties following from quantum
metamaterial being an extended quantum system, it is also a medium
with an optical response, which can be changed at will without
changing its microstructure; such media are a focus of strong
research effort \cite{Zhou2008,Gelman}.

\begin{figure}
\includegraphics{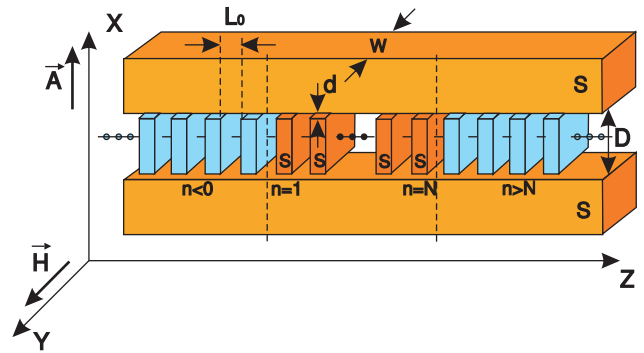}
\caption{(Color online). 1D quantum metamaterial in a
superconducting transmission line. The superconducting islands
($S$) sandwiched between the superconducting strips form charge
qubits ($1\leq n\leq N$); $D$ is the distance between the strips
($S$), $L_0$ is the distance between the qubits. In the passive
(in- and -out-) regions, with $n\leq0$ and $n>N$ respectively, the
transmission line parameters are chosen to optimize impedance
matching  with the active (metamaterial) section. The
electromagnetic pulse propagates along $z$ axis, $\vec{A}$ is the
vector potential, $\vec{H}$ is the magnetic field of the
electromagnetic wave.}\label{fig_s1}
\end{figure}

In this paper we show that the local manipulation of the quantum
states of the constituent units of a quantum metamaterial is not
the only possible way of initializing it in a desirable state. To
be specific, we show that a spatially-periodic quantum state of a
1D quantum metamaterial (i.e., quantum photonic crystal \cite{6})
can be realized by sending, from opposite directions, a pair of priming pulses through the
system. This provides an easier way to
initialize a quantum metamaterial, since it would not require the
local control of each qubit in it. Disposing of this requirement
significantly simplifies the experimental realization of quantum
metamaterials by reducing the following: the complexity of their
design, the coupling to the environment, and therefore the
decoherence due to both internal and external sources. 

We  model   a 1D quantum metamaterial using superconducting
charge qubits \cite{6}, though the main idea of our approach is
independent of the particular implementation. The (identical)
charge qubits are formed by superconducting islands sandwiched
between superconducting strips and separated from them by
tunneling barriers (see Fig.~\ref{fig_s1}). The metamaterial
occupies the central (``active'') section of the transmission line
formed by the strips; in the rest of the system (``passive'')
qubits are replaced by capacitive elements chosen to optimize the
impedance matching between the active and passive sections of the
line.

As in \cite{6}, we treat the electromagnetic field in the system
classically, while quantizing the qubit degrees of freedom. The
field propagates along the $z$ direction. Neglecting edge effects, the
magnetic field will have only a $y$-component, and the vector
potential in a section between the  $n$th and $(n+1)$st qubits can
be chosen  as $A_{xn} \hat{e}_x$. The energy of the active part of
the system can be then written as
\begin{widetext}
\begin{eqnarray}\label{s1}
    E_{SC}=\sum_{n=1}^{n=N}\Biggl[\frac{E_J}{2\omega_J^2}
    \left(\left(\frac{2\pi
    D\dot{A}_{xn}}{\Phi_0}+\dot{\varphi}_n\right)^2
    +\left(\frac{2\pi
    D\dot{A}_{xn}}{\Phi_0}-\dot{\varphi}_n\right)^2\right)+\\
    -E_J\left(\cos\left(\varphi_n+\frac{2\pi DA_{xn}}{\Phi_0}\right)+
    \cos\left(\varphi_n-\frac{2\pi
DA_{xn}}{\Phi_0}\right)\right) +
\frac{DL_0W\dot{A}^2_{xn}}{8\pi c^2}+
\frac{DL_0W}{8\pi}\left(\frac{A_{x,n+1}-A_{x,n}}{L_0}\right)^2\Biggr].\nonumber
\end{eqnarray}
\end{widetext}
Here $\varphi_n$ is the superconducting phase of the $n$th island,
$E_J=\Phi_0I_c/2\pi c$ and $\omega_J^2=2eI_c/\hbar C$ are the
Josephson energy and Josephson frequency respectively, $I_c$ is
the critical current, $C$ is the Josephson junction capacity,
$\Phi_0=hc/2e$ is the magnetic flux quantum, $D$ is the distance
between the superconducting lines, $W$ is the thickness of the
superconducting strips. The first term in Eq.~(\ref{s1}) is the
charging energy of the $n$th qubit, the second is the Josephson
energy, and the last two terms describe the electromagnetic field energy. 
The time derivative is indicated by a dot ($\dot{A}, \dot{\varphi}$). We
assume here that the qubit size is much smaller than the distance
between the neighbouring qubits and that the vector potential does not
depend on the coordinates in the area between them. Similarly, in
the passive regions
\begin{eqnarray}\label{s1-2}
    E_{NC}=\sum_{n\leq0,n>N}\Biggl[\frac{\tilde{C}D^2{\dot A}^2_{xn}}{c^2}+\frac{DL_0W\dot{A}^2_{xn}}{8\pi
    c^2}+\\\nonumber
    +\frac{DL_0W}{8\pi}\left(\frac{A_{x,n+1}-A_{x,n}}{L_0}\right)^2\Biggr].
\end{eqnarray}
Comparing the expressions for the electromagnetic field energy for
the active [Eq.~(\ref{s1})] and the passive [Eq.~(\ref{s1-2})]
regions one can conclude that the impedance matching is fulfilled
if $\tilde{C}=4e^2E_J/\hbar^2\omega_J^2$.

In order to obtain  the Hamiltonian from the full classical energy of the system, $E_{SC}+E_{NC}$, we quantize the qubit contributions by writing
\begin{eqnarray}\label{s2}
\hat{H}_n\left|n\right>=E_n\left|n\right>,\nonumber\\
    \hat{H}_n=-\frac{(\hbar\omega_J)^2}{E_J}\left(\frac{\partial}{\partial\varphi_n}\right)^2-2 E_J\cos\varphi_n  + \hat{h}_n.
    \label{eq-hhat}
\end{eqnarray}
Here the term $\hat{h}_n$ describes the external controls of the state of the $n$-th qubit; its precise form is irrelevant here. Then we restrict the operator $\hat{H}_n$ to the subspace of its two lowest
states, $|0_n\rangle$ and $|1_n\rangle$, while $E_{1,n}~-~E_{0,n} =
\hbar\varepsilon_n$.    We can assume this due to the nonlinearity of the
Josephson potential, assuming that both the
temperature and the field amplitudes are small enough to keep only
these states populated.

As follows from Eq.~(\ref{s1}), the interaction of the qubit with
the electromagnetic field is given by
\begin{equation}\label{s2_v}
    \hat V_n=2E_J(1-\cos a_n)\cos\varphi_n,
\end{equation}
where  $a_n=2\pi DA_{xn}/\Phi_0$ is the dimensionless vector
potential.

In the following, we consider the simplest case of factorizable
wave function of the metamaterial. Each qubit is described by its
own wave function $\left|\psi_n\right>$:
\begin{equation}\label{c3}
    \left|\psi_n(t)\right>=c_0(n,t)\left|0\right>e^{i\varepsilon t/2}+c_1(n,t)\left|1\right>e^{-i\varepsilon
    t/2};
\end{equation}
we omit here the subscripts in $|0\rangle$ and $|1\rangle$, since the
qubits are assumed to be identical. The Shr\"{o}dinger equation
for $\left|\psi_n\right>$ is thus reduced to a set of equations for the
coefficients $c_0$ and $c_1$:
\begin{equation}\label{c4}
i\hbar\dot{c}_{\alpha}(n,t)=\sum_{\beta=0,1}\left<\alpha\right|V_{n}(t)\left|\beta\right>c_{\beta}(n,t)e^{i\left(\omega_{\alpha}-\omega_{\beta}\right)t},
\end{equation}
where $(\alpha,\beta)=0,1$, $\omega_1=\varepsilon/2$ and
$\omega_0=-\varepsilon/2$. From Eq.~(\ref{s1}) follows the equation
for the electromagnetic field in the active region $1 \leq n\leq
N$:
\begin{eqnarray}\label{s3}
    \ddot{a}_n-\upsilon^2(a_{n+1}+a_{n-1}-2a_n)+\\\nonumber r\sin
    a_n\left<\psi_n\right|\cos\varphi_n\left|\psi_n\right>=0,
\end{eqnarray}
where
\begin{equation}
\upsilon^2=\frac{rW\Phi_0^2}{32\pi^3L_0DE_J},\ \ \
r=\omega_J^2\left(1+\frac{\Phi_0^2L_0W\omega_J^2}{32\pi^3c^2DE_J}\right)^{-1}.
    \label{eq:xyz}
\end{equation}
In the passive regions we have instead
\begin{equation}\label{s3-1}
     \ddot{a}_n - u^2(a_{n+1}+a_{n-1}-2a_n)=0.
\end{equation}
With the proper choice of parameters in the passive region one can obtain $u = \upsilon$.

For an analytical treatment, we make additional plausible
assumptions. First, the electromagnetic field must be weak,
$a_n\ll 1$ (otherwise the two-level approximation for qubits may
not be valid); therefore
\begin{equation}\label{c5}
    \hat
    V_n \approx E_Ja_n^2\cos{\varphi_n},
\end{equation}
and $\sin{a_n}\approx a_n$. Second, the field wavelength under a
realistic choice of parameters  greatly exceeds the interqubit
distance and therefore allows to introduce a continuous variable
$z$ instead of the discrete label $n$ (Ref.~[\onlinecite{6}]). Then
Eq.~(\ref{s3}) is transformed into:
\begin{equation}\label{s4}
    \frac{\partial^2a}{\partial t^2}-\tilde{\upsilon}^2\frac{\partial^2a}{\partial
    z^2}+\chi(z,t) a=0,
\end{equation}
where $\tilde{\upsilon}=\upsilon L_0$, $a \equiv a(z,t)$, and the
``susceptibility''
\begin{equation}\label{s4-1}
\chi(z,t)=r\left<\psi(z,t)\right|\cos\varphi\left|\psi(z,t)\right>
\end{equation}
is determined by the quantum state of the metamaterial.

In agreement with our weak-field assumption, we analyze
Eq.~(\ref{s4}) in a quasilinear approximation. These analytical
results will help illuminate the numerical solutions of
Eqs.~(\ref{c4},\ref{s3},\ref{s3-1}) for the following choice of
parameters: $L_0=5\cdot10^{-4}\ m$, $D \sim W \sim 10^{-5}\ m$,
critical current $I_c=4\cdot 10^{-7}\ A$, $\varepsilon/2\pi \simeq
5\cdot10^{10}\ s^{-1}$, and $E_J/\hbar\omega_J=4$, so,
$\tilde{\upsilon}\simeq c$.

As shown in Ref.[\onlinecite{6}], the transmission properties of a 1D
metamaterial with a spatially-periodic quantum state, $|\psi(z +
L_m,t=0)\rangle = |\psi(z,t=0)\rangle$, are those of a generalized
photonic crystal. We therefore investigate the possibility of
initializing the metamaterial without applying local controls to
the constituent qubits. Specifically, let us initialize the
metamaterial in the ground state (which can be done by, e.g.,
cooling) and send two priming pulses through the active region of
the system in opposite directions. One would expect that due to
the interference of these pulses and their action on the qubits, a
spatially-periodic state of the quantum metamaterial would arise.
For superconducting qubits the decoherence times are currently in
the range 10-100 $\mu$s and the interlevel distance is $\sim 10$
GHz; the transition time for the pulse across a 1000-unit
metamaterial (of length $\sim 10^2$ cm) is $\sim 3 \cdot 10^{-9}$
s. Therefore, in the following, we can neglect decoherence
effects.

The backaction of the electromagnetic field on the qubits is
quadratic in the field amplitude (Eq.~\ref{c5}), therefore the
basic frequency of the priming pulses should satisfy the condition
$2\omega\approx\varepsilon$ (see, e.g.,~\cite{9,eberly,6}). We
assume that the spatial width of the pulse satisfies
\begin{equation}\label{c2-0}
l \gg {2\pi \tilde{\upsilon}/\omega} \approx \lambda,
\end{equation}
which ensures that each qubit in a superposition state undergoes
many cycles of quantum beats while the pulse propagates past it.
Here $\lambda$ is the wavelength of the priming pulses. In this
case, the metamaterial state evolution in the presence of the
field can be investigated in the resonance approximation\cite{9}. So, Eqs.~(\ref{c4}) become:
\begin{eqnarray}\label{c6}
    i\hbar\dot{c}_0(z,t)=a^2(z,t)\left(d_{00}c_0+d_{01}c_1e^{-i\varepsilon
    t}\right),\\
    i\hbar\dot{c}_1(z,t)=a^2(z,t)\left(d_{10}c_0e^{i\varepsilon
    t}+d_{11}c_1\right),\nonumber
\end{eqnarray}
where
$d_{\alpha\beta}=E_J\left<\alpha\right|\cos{\varphi_n}\left|\beta\right>$
and $z=nL_0$. In the case of sufficiently strong pulses the
solution of Eq.(\ref{s4}) may be written as
\begin{eqnarray}\label{c7}
    a^{(1)}(z,t)=\exp\left[-\frac{\left[z-(\omega/k)t\right]^2}{l^2}\right]\left(Ae^{i(kz-\omega t)}+c.c.\right),\nonumber\\
    \\
a^{(2)}(z,t)=\exp\left[-\frac{\left[z+(\omega/k)
t\right]^2}{l^2}\right]\left(Ae^{i(kz+\omega
t+\phi_0)}+c.c.\right),\nonumber
\end{eqnarray}
where $A$ is the amplitude of the pulses (the pulses have equal
amplitudes), and $\phi_0$ is the initial phase. In order for our
treatment of the metamaterial as a continuous medium to be
consistent, the wave vectors should satisfy the condition $\lambda
= 2\pi/k \gg L_0$, that is
\begin{equation}
L_0  \ll 2\pi\tilde{\upsilon}/\omega.
    \label{eq:L0condition}
\end{equation}
For $\tilde{\upsilon} \approx c$ and $\varepsilon/2\pi \approx 5\cdot
10^{10}\ {\rm s}^{-1}$, this yields $L_0 \ll 1\ {\rm cm}$, which
is a feasible requirement.

Let all the qubits be initialized at $t=0$ in the ground state
[$c_0(z,0)=1, c_1(z,0)=0$], then, at later times, the
quasi-monochromatic approximation [Eq.~(\ref{c2-0})] gives
\begin{equation}\label{c9}
    |c_1(z,t)|=\frac{|\Omega (z)|\sin\left(\sqrt{|\Omega (z)|^2+\gamma (z)^2/4}t\right)}{\sqrt{|\Omega (z)|^2+\gamma (z)^2/4}},
\end{equation}
\begin{equation}\label{c8}
    \gamma (z)=\Delta+4A^2\frac{d_{00}-d_{11}}{\hbar}\left[\cos(2kz+\phi_0)+1\right],
\end{equation}
where $\Delta=2\omega-\varepsilon$ is the detuning from the
resonance. The local Rabi frequency is
\begin{equation}\label{c10}
    |\Omega(z)|=\frac{2|d_{01}|A^2}{\hbar}\left[\cos(2kz+\phi_0)+1\right].
\end{equation}
The periodicity of $\gamma (z)$ and $\Omega(z)$
[Eqs.~\ref{c8},~\ref{c10}] implies a spatially-periodic
probability of exciting the qubit with period $\lambda/2$.

\begin{figure}
\includegraphics[width=3.5in]{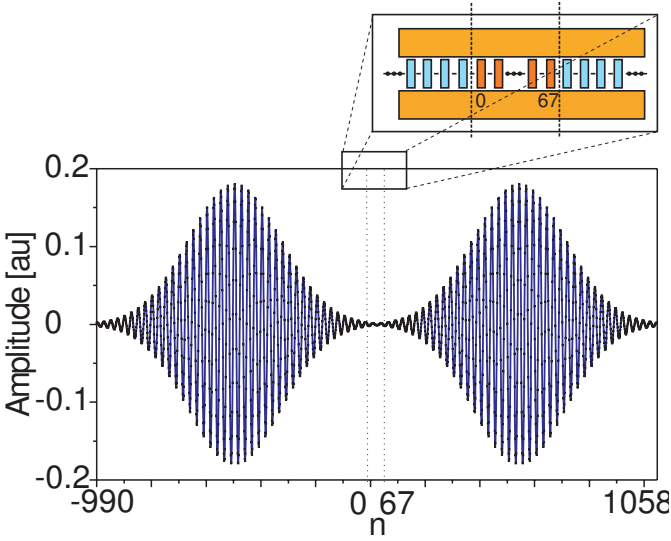}
\caption{(Color online). The priming or initializing pulses
located in the passive regions of the waveguide line at the
initial moment of time. The active region is in the center of the
line between the two vertical dotted lines. The pulse parameters
are $\tilde{\upsilon}=c$,
   $\omega = \varepsilon/2$, $\Delta=0.18 \varepsilon$,
 $k=(1/25)\cdot2\pi/L_0$. The pulse width is $l=240 L_0$, and the pulse
amplitude $A=0.18$. The matrix elements are
$d_{00}=0.4 \hbar\varepsilon$, $d_{11}=3.6 \hbar\varepsilon$, and
$d_{01}=0.2 \hbar\varepsilon$. The total length of the system
simulated here is $\tilde{L}=2048 L_0$.}\label{pic2}
\end{figure}

\begin{figure}
  \includegraphics[width=3.5in]{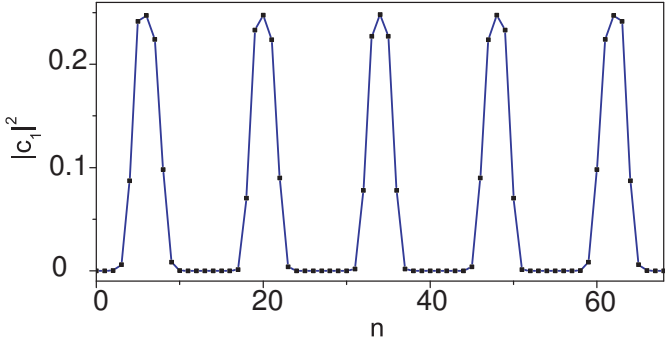}
\caption{(Color online). Periodically-modulated population of the
excited levels of qubits after passing the priming or initializing
pulses through the system. The period of modulation, $13 L_0
\approx \lambda/2$, shows that this is due to the interference
between these pulses. The slight aperiodicity is caused by the
finite width of the pulse.}\label{pic3}
\end{figure}

Figures \ref{pic2} and \ref{pic3} show the results of a numerical
simulation of such a process from Eqs.~(\ref{s3},\ref{s3-1}). Two
priming pulses at $t=0$ are located in the passive regions of the
waveguide (Fig.~\ref{pic2}). After passing through the active
region, they produce a periodically-modulated population of the
levels of the qubits (Fig.~\ref{pic3}). A small distortion of the
periodicity is due to the fact that the waves are
non-monochromatic and can be improved by using wider pulses. For
the system parameters chosen, the period of modulation in
Fig.~\ref{pic3} is approximately equal to $13L_0$, in good
agreement with the modulation period  $\pi/k = 12.5 L_0$ following
from the approximate solution Eq.~(\ref{c10}). The modulation
amplitude can also be controlled by the pulse amplitude and the
pulse width [cf. Eqs.~(\ref{c9},~\ref{c10})].


The periodic spatial modulation of the wave function
$|\psi(z)\rangle$, created by the priming pulses, will affect the
propagation of the subsequent probe pulse through the function
$\chi(z)$ [Eq. (\ref{s4-1})], which enters the wave equation
(\ref{s4}):
\begin{eqnarray} \chi(z,t)=\chi_0+\tilde{\chi}\left[1+\cos({2\pi
z}/{L_m})\right]+\nonumber\\
    +2\frac{r}{E_J}Re[d_{01}c_0^*(z)c_1(z)e^{-i\varepsilon t}],
\label{eq:z1}
\end{eqnarray}
where $L_m$ is the modulation period, and $\chi_0$ depends on the
quantum states of the qubits with minimal $|c_{1}|$:
\begin{equation}
\chi_0=\frac{r}{E_J}\left(d_{00}|c_{0,\textrm{min}}|^2+d_{11}|c_{1,\textrm{min}}|^2\right).
\label{eq:z2}
\end{equation}

For the case shown in Fig.~\ref{pic3}, $\chi_0=(r/E_J)d_{00}$,
since the qubits at positions such that $\cos(2kz+\phi_0)=-1$
(i.e., where the local Rabi frequency is zero)  remain in the
ground state, and thus $|c_{1,\textrm{min}}|=0$. The quantity
$\tilde{\chi}$ describes the  excitation  of qubits at positions
where $\cos(2kz+\phi_0)=1$, and $|c_1|=|c_{1,\textrm{max}}|$:
\begin{equation}
\tilde{\chi}=r\left(d_{11}-d_{00}\right)\left(|c_{1,\textrm{max}}|^2-|c_{1,\textrm{min}}|^2\right)/2E_J.
\label{eq:z3}
\end{equation}
For the case of Fig.~\ref{pic3} we have
$\tilde{\chi}=r\left(d_{11}-d_{00}\right)|c_{1,\textrm{max}}|^2/2E_J.$

If
$$\tilde{\chi}\gg |\int_{-L_m/2}^{L_m/2}{[(r/E_J)d_{01}c_0^*(z)c_1(z)e^{i2\pi
z/L_m}]dz}|,$$ then one can neglect the third term in
Eq.~(\ref{eq:z1}) to derive the dispersion equation and thus
consider $\chi$ as independent of time. So, $\chi$ in
Eq.~(\ref{eq:z1}) can be approximated as
\begin{equation}\label{c00}
    \chi(z)=\chi_0+\tilde{\chi}\left[1+\cos({2\pi
    z}/{L_m})\right].
\end{equation}

\begin{figure}
\includegraphics{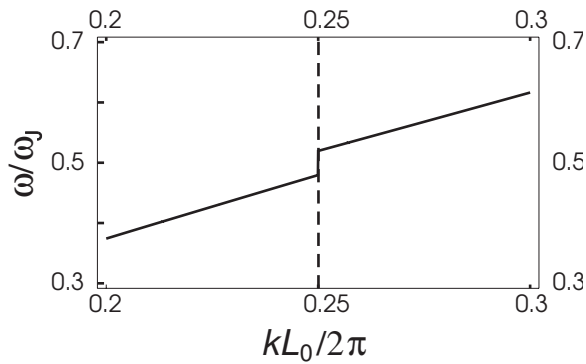}
\caption{Normalized dispersion relation $\omega(k)$ for the quantum photonic
crystal. Due to the space periodicity of $\chi(z)$, gaps open in
the spectrum. Their positions and magnitude depend on   $L_m$ and
$\tilde{\chi}$. This  allows to control the parameters of the
quantum photonic crystal by varying the quantum states of the
qubits. The graph is plotted for $L_m = 25 L_0/2$,
$d_{00} = 0.4 \hbar\varepsilon$, $d_{11} = 3.6 \hbar\varepsilon$,
$d_{01} = 0.2 \hbar\varepsilon$, and $\tilde{\upsilon} = c$. The qubit
population is the same as in
 Fig.~\ref{pic3}.}\label{pic1}
\end{figure}

We seek the solution of Eq.(\ref{s4}) with the above $\chi(z)$ for
small wave amplitude in the form of a Bloch wave~\cite{davydov}:
\begin{equation}\label{c1b}
    a_k(z,t)=u_k(z)\exp\left[i(kz-\omega t)\right],
\end{equation}
where $u_k(z)$ is a periodic function with period $L_m$ [i.e.,
$u_k(z)=u_k(z+L_m)$]. The dispersion equation is readily found:
\begin{equation}\label{c2}
    \omega_k^2-\tilde{\upsilon}^2k^2=\pm |W_k|+W_{k=0},
\end{equation}
where
\begin{equation}\label{c2a}
W_k=\frac{1}{L_m}\int_{-L_m/2}^{+L_m/2}\chi(z)\exp\left(i2kz\right)dz,\
\ k=\frac{\pi
    n}{L_m}.
\end{equation}
It is clear from Eq.~(\ref{c2}) that a gap opens in the frequency
spectrum for each $k=\pi n/L_m$ (if the corresponding Fourier
component of $\chi(z)$ is different from zero, see
Fig.\ref{pic1}):
    \begin{equation}\label{c2b}
\delta\omega_n\approx\frac{W_{k_n}}{\sqrt{\left(\pi\tilde{\upsilon}n/L_m\right)^2+W_0}}\
,
    \ \ |W_{k_n}|\ll \left(\frac{\pi\tilde{\upsilon}n}{L_m}\right)^2+W_0.\nonumber
\end{equation}
The gap appears for $k_{n=1}=\pi/L_m$, close to the frequency
$\omega_1=\sqrt{\left(\tilde{\upsilon} k_1\right)^2+W_0}$.
According to Eq.~(\ref{c2a}), $W_{k_{n=1}}=\tilde{\chi}/2$,
$W_0=\chi_0+\tilde{\chi}$, and the  gap equals
\begin{equation}\label{c2c}
    \delta\omega_1\approx\tilde{\chi}/\sqrt{\left(\pi\tilde{\upsilon}/L_m\right)^2+W_0}.
\end{equation}

\begin{figure}
  \includegraphics{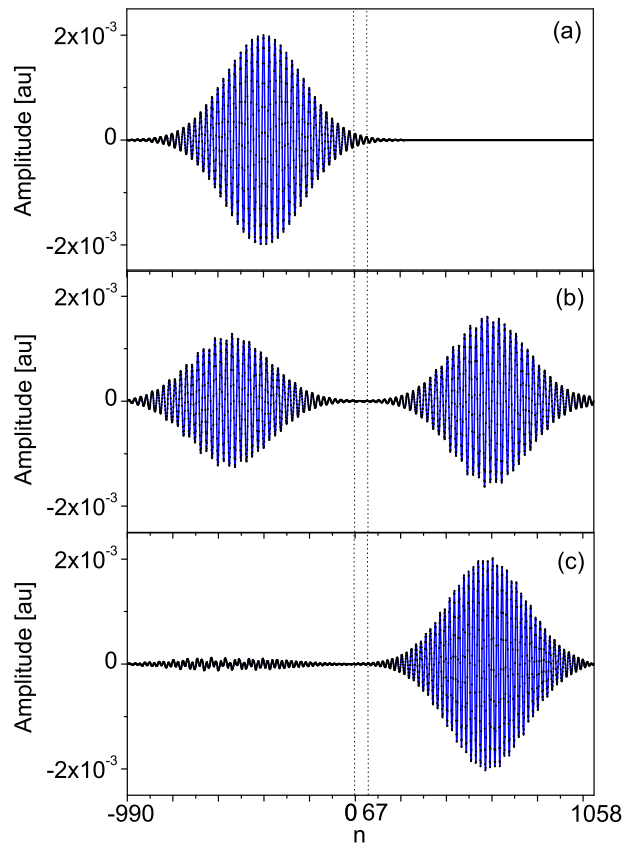}
\caption{(Color online). Probe pulse propagation through a
spatially-periodic quantum metamaterial prepared by the priming
pulses. The vertical dotted lines indicate the location of the
chain of qubits (i.e., the actual metamaterial). The metamaterial
parameters are the same as in Fig.~\ref{pic2};
$d_{00} = 0.4 \hbar\varepsilon$, $d_{11} = 3.6 \hbar\varepsilon$,
$d_{01} = 0.2 \hbar\varepsilon$; the qubit population is the same as in
Fig.~\ref{pic3}.
  (a) The probe pulse at $t=0$. The pulse parameters are: amplitude  $A=2\cdot 10^{-3}$, width $l=240L_0$,
carrier frequency $\omega = \varepsilon/2$, and velocity
  $\tilde{\upsilon}=c$.  (b) Field
distribution at $t=500 \varepsilon^{-1}$. The partial pulse
transmission is due to the finite pulse width. (c) Field
distribution at $t=500 \varepsilon^{-1}$ for a probe pulse with  
frequency above the bandgap. The pulse parameters at $t=0$ are the
same as above, except the carrier frequency now is
$\omega=0.6 \varepsilon$.}\label{fig4}
\end{figure}

In Fig.~\ref{fig4} the propagation of a probe pulse through a
metamaterial with periodic  $\chi(z)$ is shown. One can see that
while the pulse with the carrying frequency inside the bandgap
undergoes significant reflection [Fig.~\ref{fig4}(a)], this is not
the case for a pulse with a frequency above the bandgap
[Fig.~\ref{fig4}(b)].

In conclusion, we have shown that a 1D quantum metamaterial
comprised of superconducting charge qubits in a transmission line
can be initialized in a spatially-periodic state without
excercising local control of the quantum state of individual
qubits, by simultaneously passing through it two priming pulses in
opposite directions. The modulation period is close to half the
priming pulse wavelength, $\lambda/2$.  The subsequent probe pulse
propagation through the resulting periodic structure was shown to
be affected by the arising bandgaps.

It is possible that such 1D quantum metamaterials could be realized
using the microwave transmission line that was recently 
experimentally investigated in \cite{jerger}, where a set of
resonators coupled with qubits was connected with this line. If
the transmission line is long enough, we can create a more complex
picture of the qubit state distribution by varying the shape of
the pump pulses.

\begin{acknowledgments}
This work was funded in part by the Russian Ministry of Education and Science through the programs (No. 07.514.11.4147,14.B37.21.0079), and the Russian Foundation for Basic Research (Grants No. 12-07-00546,12-07-31144). FN was partially
supported by   ARO,   JSPS-RFBR contract No. 12-02-92100, Grant-in-Aid for Scientific Research
(S), MEXT Kakenhi on Quantum Cybernetics, and the JSPS via its
FIRST program. SS and AZ acknowledge that this publication was
made possible through the support of a grant from the John
Templeton Foundation.
\end{acknowledgments}

\end{document}